# A Precise Determination of the Running Coupling in the SU(3) Yang-Mills Theory

Martin Lüscher and Rainer Sommer

Deutsches Elektronen-Synchrotron DESY
Notkestrasse 85, D-22603 Hamburg, Germany

Peter Weisz

Max Planck Institut für Physik
Föhringerring 6, D-80805 München, Germany

Ulli Wolff

CERN, Theory Division, CH-1211 Genève 23, Switzerland

## Abstract

A non-perturbative finite-size scaling technique is used to study the evolution of the running coupling (in a certain adapted scheme) in the SU(3) Yang-Mills theory. At low energies contact is made with the fundamental dynamical scales, such as the string tension $K$, while at larger energies the coupling is shown to evolve according to perturbation theory. In that regime the coupling in the $\overline{\mathrm{MS}}$ scheme of dimensional regularization is obtained with an estimated total error of a few percent.

## 1. Introduction

Asymptotically free theories, such as QCD, behave very differently at low and high energies. While there is no obvious physical relation between, say, pion-nucleon scattering phases and jet production rates at $e^+e^-$ colliders, the underlying theory describes both phenomena and thus provides a link between the two.

An interesting observation in this context is that the running coupling at high energies may in principle be computed once the parameters of the theory are fixed at low energies [1]. This computation is difficult, because one refers to the non-perturbative low-energy properties of the theory. Moreover, one must be able to cover a substantial range of energies to make contact with the scaling regime, where the running coupling is logarithmically decreasing according to the perturbative renormalization group.

In this paper the problem is solved for the SU(3) Yang-Mills theory. The computation is set up in the framework of lattice gauge theories and relies on numerical simulations. To step up the energy scale, a recursive finite-size scaling technique is employed. The method has previously been applied to the SU(2) theory [3,4] and we shall here assume that the reader is familiar with this work (for other approaches see refs.[6–11]).

There are two new features in our computations. The first is that we now use a 1-loop improved lattice action, which leads to significantly reduced lattice spacing effects. This makes the extrapolation of the lattice data to the continuum limit safer.

The other change concerns the choice of reference scale at low energies. In the SU(2) theory all physical momenta were given in units of the string tension $K$. From a technical point of view this quantity is somewhat problematic, because in practice it is determined by extrapolating the heavy quark force $F(r)$ from distances $r$ less than 1 fm to large distances. Depending on which analytical form is assumed for the extrapolation, one can obtain quite different results. In other words, the string tension is affected by a systematic error which is not easy to control.

An alternative reference scale is the distance $r_0$ at which

$$r_0^2 F(r_0) = 1.65 \tag{1.1}$$

The merits of this definition have been discussed in ref.[5]. The most important points to note are the following.

1. $r_0$ is well-defined in $\mathrm{SU}(N)$ gauge theories with or without matter fields.

2. From the non-relativistic charmonium model one estimates that $r_0$ is approximately equal to 0.5 fm in nature [the number on the right hand side of eq.(1.1) has been chosen so as to achieve this]. The exact relation of $r_0$ to other scales in QCD, such as the pion decay constant, is presently not known.

3. In lattice gauge theories $r_0$ is comparatively easy to compute through numerical simulation, because one only requires the static quark force in an accessible range of distances. No extrapolation is needed here.

4. In the SU(3) Yang-Mills theory one finds

$$r_0\sqrt{K} = 1.22(8), \tag{1.2}$$

where the error includes an estimate of the systematic uncertainty on the string tension mentioned above (cf. sect. 5).

Neither the distance $r_0$ nor the string tension are experimentally measurable, but $r_0$ has several conceptual and technical advantages so that we decided to express our final results in units of this scale.

## 2. Lattice theory

The study reported in this paper is a straightforward extension of our earlier work on the SU(2) Yang-Mills theory [2–4]. We assume the reader is familiar with our computational strategy. Here we only introduce the necessary notations.

We choose to set up the theory on a hyper-cubic euclidean lattice with spacing $a$ and size $L\times L\times L\times L$. The possible values of the time coordinate $x^0$ of a lattice point $x$ are $x^0 = 0, a, 2a, \ldots, L$ ($L$ is taken to be an integer multiple of $a$). The spatial sublattices at fixed times are thought to be wrapped on a torus, i.e. we assume periodic boundary conditions in these directions.

A gauge field $U$ on the lattice is an assignment of a matrix $U(x,\mu)\in\mathrm{SU}(3)$ to every pair $(x, x+a\hat{\mu})$ of nearest neighbour lattice points ($\hat{\mu}$ denotes the unit vector in the $\mu$–direction and $\mu = 0,1,2,3$). At the top and bottom of the lattice, the link variables are required to satisfy inhomogeneous Dirichlet boundary conditions,

$$U(x,k)|_{x^0=0} = W(\mathbf{x},k), \qquad U(x,k)|_{x^0=L} = W'(\mathbf{x},k), \tag{2.1}$$

for all $k = 1,2,3$, where $W$ and $W'$ are prescribed spatial gauge fields. They will be set to some particular values below.

The partition function of the system,

$$\mathcal{Z} = \int \mathrm{D}[U]\,\mathrm{e}^{-S[U]}, \qquad \mathrm{D}[U] = \prod_{x,\mu} \mathrm{d}U(x,\mu), \tag{2.2}$$

involves an integration over all fields $U$ with fixed boundary values $W$ and $W'$. For the action $S[U]$ we take

$$S[U] = \frac{1}{g_0^2}\sum_p w(p)\,\mathrm{tr}\left\{1 - U(p)\right\}, \tag{2.3}$$

with $g_0$ being the bare coupling. The sum in eq.(2.3) runs over all *oriented* plaquettes $p$ on the lattice and $U(p)$ denotes the parallel transporter around $p$. The weights $w(p)$ depend on whether one desires to work with the simple Wilson action or an improved action. In the first instance $w(p)$ is equal to 1 except for the spatial plaquettes at $x^0 = 0$ and $x^0 = L$ which are given the weight $\frac{1}{2}$.

As discussed in sect. 4 of ref.[2], the leading cutoff effects in the Wilson theory are of order $a$. They can be cancelled by adjusting the weights of the plaquettes at the boundary of the lattice, i.e. one sets

$$w(p) = \begin{cases} \frac{1}{2}c_s(g_0) & \text{if } p \text{ is a spatial plaquette at } x^0 = 0 \text{ or } x^0 = L, \\ c_t(g_0) & \text{if } p \text{ is a time-like plaquette attached to a} \\ & \text{boundary plane,} \end{cases} \tag{2.4}$$

and $w(p) = 1$ in all other cases. With appropriately chosen coefficients $c_s(g_0)$ and $c_t(g_0)$, the cutoff effects are then reduced to order $a^2$.

We shall exclusively be concerned with constant abelian boundary fields $W$ and $W'$. Since the contribution of the spatial boundary plaquettes to the action vanishes in this case, we do not need to know the value of $c_s(g_0)$. For the other improvement coefficient we use the 1-loop formula

$$c_t(g_0) = 1 + c_t^{(1)} g_0^2, \qquad c_t^{(1)} = -0.08900(5). \tag{2.5}$$

This removes all cutoff effects of order $a$ at the tree and the 1-loop level of perturbation theory.

## 3. Running coupling

From the partition function $\mathcal{Z}$ (which is also referred to as the Schrödinger functional) a running coupling may be defined by differentiating with respect to the boundary values [2]. To obtain a unique coupling, the boundary fields and the direction of differentiation must be specified. There are no deep theoretical reasons for the choices made below. They are, however, based on practical considerations, such as the requirement of mild cutoff effects and calculability.

*3.1 Boundary values and induced background field*

As in the SU(2) theory we choose the boundary values $W$ and $W'$ to be constant diagonal matrices. More precisely we set

$$W(\mathbf{x},k)=\exp\{aC_k\} \quad \text{and} \quad W'(\mathbf{x},k)=\exp\{aC'_k\}, \tag{3.1}$$

where

$$C_k=\frac{i}{L}\begin{pmatrix}\phi_1 & 0 & 0\\ 0 & \phi_2 & 0\\ 0 & 0 & \phi_3\end{pmatrix} \quad \text{and} \quad C'_k=\frac{i}{L}\begin{pmatrix}\phi'_1 & 0 & 0\\ 0 & \phi'_2 & 0\\ 0 & 0 & \phi'_3\end{pmatrix}. \tag{3.2}$$

The angles $\phi_\alpha$ and $\phi'_\alpha$ must be real and they should add up to zero to guarantee that the boundary link variables are elements of SU(3).

The classical field

$$V(x,\mu)=\exp\{aB_\mu(x)\}, \tag{3.3}$$

with

$$B_0=0, \qquad B_k=\left[x^0C'_k+(L-x^0)C_k\right]/L, \tag{3.4}$$

is a solution of the lattice field equations with the required boundary values. $V$ is referred to as the induced background field. A theorem proved in ref.[2]

asserts that $V$ is absolutely stable if $L/a \geq 5$ and if the vectors $(\phi_1, \phi_2, \phi_3)$ and $(\phi_1', \phi_2', \phi_3')$ are in a certain bounded region, the *fundamental domain.*

Stability ensures that at small couplings $g_0$, the Schrödinger functional is dominated by the field configurations in a neighbourhood of the induced background field. The perturbation expansion then amounts to a saddle point expansion about $V$.

*3.2 Geometry of the fundamental domain*

The fundamental domain consists of all vectors $(\phi_1, \phi_2, \phi_3)$ satisfying

$$\phi_1 + \phi_2 + \phi_3 = 0, \qquad \phi_1 < \phi_2 < \phi_3, \qquad \phi_3 - \phi_1 < 2\pi. \tag{3.5}$$

This is a 2-dimensional simplex, which can be described in a more symmetrical way as follows. Let

$$\mathbf{e}_1 = (1,0), \qquad \mathbf{e}_2 = \tfrac{1}{2}(-1, \sqrt{3}), \qquad \mathbf{e}_3 = \tfrac{1}{2}(-1, -\sqrt{3}) \tag{3.6}$$

be the corners of an equilateral triangle in the plane. To any vector $(\phi_1, \phi_2, \phi_3)$ in the fundamental domain, we may associate a point $\mathbf{v}$ in the plane through

$$\mathbf{v} = \tfrac{2}{3}\left(\phi_1\mathbf{e}_1 + \phi_2\mathbf{e}_2 + \phi_3\mathbf{e}_3\right). \tag{3.7}$$

This relationship is one-to-one, since

$$\phi_\alpha = \mathbf{v} \cdot \mathbf{e}_\alpha \tag{3.8}$$

for all $\alpha = 1, 2, 3$. The image of the mapping is the interior of an equilateral triangle with corners

$$\mathbf{a} = \mathbf{0}, \qquad \mathbf{b} = -\frac{4\pi}{3}\mathbf{e}_1, \qquad \mathbf{c} = \frac{4\pi}{3}\mathbf{e}_3 \tag{3.9}$$

(see fig. 1). In other words, any point $\mathbf{v}$ in this area determines a vector $(\phi_1, \phi_2, \phi_3)$ in the fundamental domain and vice versa.

The triangle shown in fig. 1 is mapped onto itself under reflections at the dashed lines. A reflection at the line passing through the point $\mathbf{c}$ corresponds

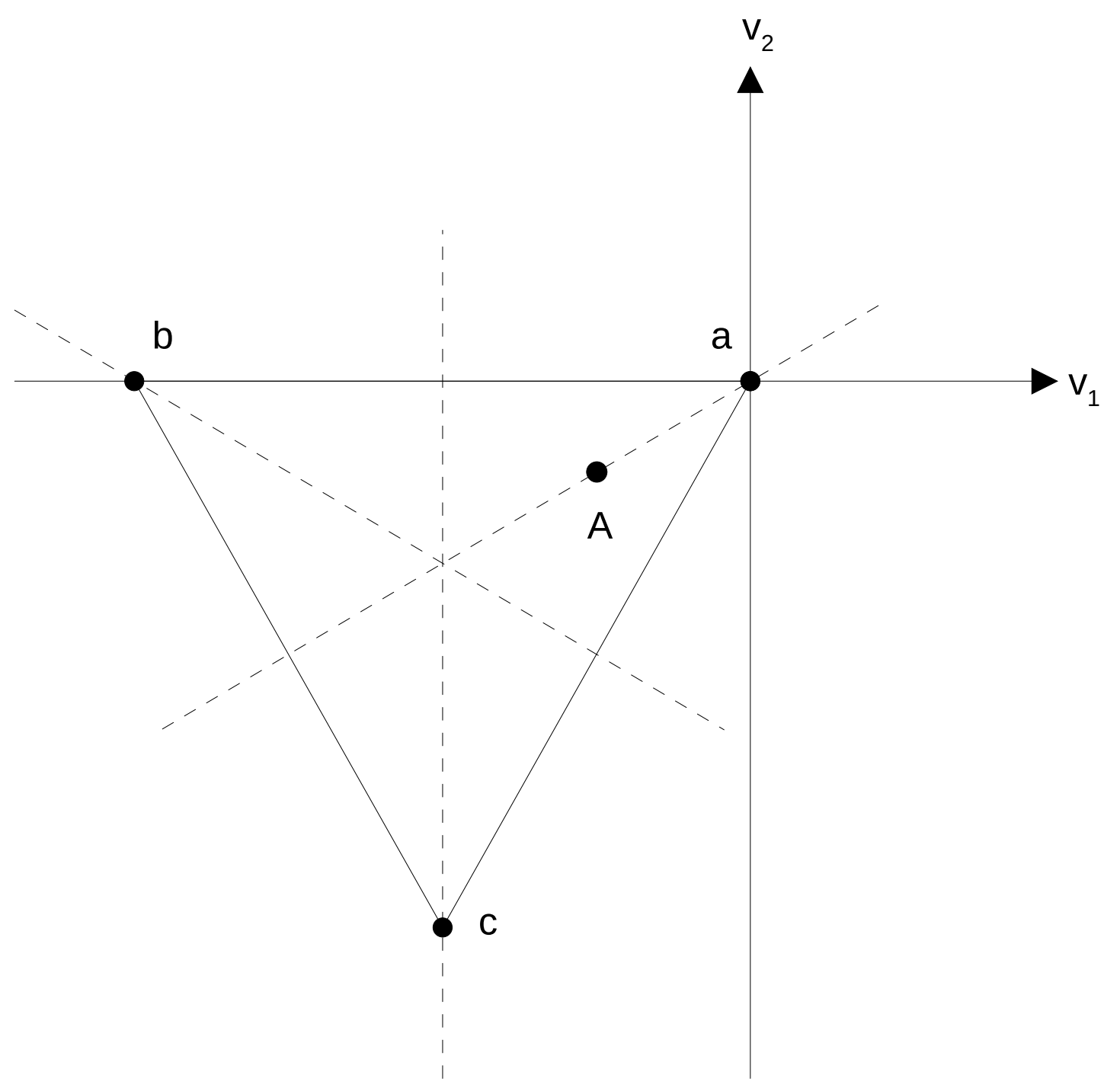

Fig. 1. Fundamental domain in the $\mathbf{v}$–plane (area enclosed by the triangle $\mathbf{abc}$). The point $\mathbf{A}$ corresponds to the angles $(\phi_1, \phi_2, \phi_3)$ chosen for the definition of the running coupling.

to a transformation $\phi_\alpha \mapsto \tilde{\phi}_\alpha$, where

$$\begin{aligned} \tilde{\phi}_1 &= -\phi_1 - \frac{4\pi}{3}, \\ \tilde{\phi}_2 &= -\phi_3 + \frac{2\pi}{3}, \\ \tilde{\phi}_3 &= -\phi_2 + \frac{2\pi}{3}. \end{aligned} \tag{3.10}$$

If we define the boundary field $W$ through eqs.(3.1),(3.2), the replacement of $\phi_\alpha$ by $\tilde{\phi}_\alpha$ amounts to a charge conjugation followed by a central conjugation (a gauge transformation which is periodic up to a central phase factor [12]). The transformation (3.10) thus corresponds to a discrete symmetry of the Schrödinger functional and so do the other reflections at the dashed lines in fig. 1.

*3.3 Definition of the running coupling*

We now choose a 1-parameter curve of vectors $(\phi_1, \phi_2, \phi_3)$ and $(\phi'_1, \phi'_2, \phi'_3)$ in the fundamental domain. The effective action

$$\Gamma = -\ln \mathcal{Z} \tag{3.11}$$

then becomes a function of the curve parameter $\eta$ and a renormalized coupling $\bar{g}^2$ may be defined by differentiating $\Gamma$ with respect to $\eta$.

The curve of angles $(\phi_1, \phi_2, \phi_3)$ that we decided to take is given by

$$\begin{aligned} \phi_1 &= \eta\, w_1 - \frac{\pi}{3}, & w_1 &= 1, \\ \phi_2 &= \eta\, w_2, & w_2 &= -\frac{1}{2} + \nu, \\ \phi_3 &= \eta\, w_3 + \frac{\pi}{3}, & w_3 &= -\frac{1}{2} - \nu. \end{aligned} \tag{3.12}$$

This is a straight line which passes through the point $\mathbf{A}$ at $\eta = 0$ and so is contained in the fundamental domain for small values of $\eta$ (see fig. 1). The direction $(w_1, w_2, w_3)$ of the line may be adjusted by changing the parameter $\nu$.

The other angles $\phi'_\alpha$ are chosen to be related to $\phi_\alpha$ through the symmetry (3.10), i.e. we set $\phi'_\alpha = \tilde{\phi}_\alpha$. The partition function is then invariant under a discrete symmetry transformation of the gauge field, which is a combination of a time reflection, a charge conjugation and a central conjugation. In particular, the boundaries at $x^0 = 0$ and $x^0 = L$ are treated equally.

We can now define a renormalized coupling $\bar{g}^2$ through

$$\left.\frac{\partial \Gamma}{\partial \eta}\right|_{\eta=\nu=0} = \frac{k}{\bar{g}^2}, \tag{3.13}$$

where

$$k = 12(L/a)^2[\sin(\theta) + \sin(2\theta)], \qquad \theta = \tfrac{1}{3}\pi(a/L)^2. \tag{3.14}$$

This choice of proportionality constant guarantees that $\bar{g}^2$ will be equal to $g_0^2$ to lowest order of perturbation theory, for all values of the lattice spacing. The reason for setting $\nu = 0$ is that in numerical simulations of the Schrödinger functional the statistical errors on the coupling turn out to be minimal in this case. Note that the box size $L$ is the only external scale appearing in the definition (3.13), i.e. $\bar{g}^2$ is a running coupling defined at distance $L$.

For general $\nu$ we have

$$\left.\frac{\partial \Gamma}{\partial \eta}\right|_{\eta=0} = k\left\{\frac{1}{\bar{g}^2} - \nu\bar{v}\right\}, \tag{3.15}$$

where $\bar{v}$ is another renormalized quantity. A moment of thought reveals that $\bar{v}$ is independent of $\nu$. Our interest in this quantity arises from the fact that it can be computed with little extra work and that it may be used to test the universality of the Schrödinger functional.

*3.4 Relation to other schemes*

In the continuum limit, and at sufficiently high energies (small $L$), different running couplings may be related in perturbation theory. The coupling defined above can be computed to 1-loop order using the techniques described in ref.[2]. Since there is no new element involved here, we only quote the results of our calculations.

Let us define

$$\alpha(q) = \frac{\bar{g}^2(L)}{4\pi}, \qquad q = 1/L. \tag{3.16}$$

The connection between the $\overline{\mathrm{MS}}$ scheme of dimensional regularization [13] and our finite volume scheme then is

$$\alpha_{\overline{\mathrm{MS}}} = \alpha + k_1\alpha^2 + \ldots, \qquad k_1 = 1.25563(4), \tag{3.17}$$

where both couplings are evaluated at the same momentum $q$.

Another coupling in infinite volume,

$$\alpha_{\mathrm{q\bar{q}}}(q) = \tfrac{3}{4}r^2F(r), \qquad q = 1/r, \tag{3.18}$$

is obtained from the force $F(r)$ between static quarks at distance $r$. Combining the expansion above with the 1-loop results of refs.[14,15], one finds

$$\alpha_{\mathrm{q\bar{q}}} = \alpha + h_1\alpha^2 + \ldots, \qquad h_1 = 1.33776(4), \tag{3.19}$$

i.e. to this order there is practically no difference between $\alpha_{\mathrm{q\bar{q}}}$ and $\alpha_{\overline{\mathrm{MS}}}$.

We mention in passing that $\bar{v}$ vanishes to lowest order of perturbation theory. The leading term in the continuum limit is

$$\bar{v} = 0.0694603(1), \tag{3.20}$$

while a determination of the next order correction (a term proportional to $\alpha$) would require a 2-loop calculation.

*3.5 Renormalization group*

We again assume that the continuum limit has been taken and define the Callan-Symanzik $\beta$–function through

$$\beta(\bar{g}) = -L\frac{\partial \bar{g}}{\partial L}. \tag{3.21}$$

From eq.(3.17) and the known perturbation expansion of the $\beta$–function in the $\overline{\mathrm{MS}}$ scheme, we infer that

$$\beta(g) \underset{g\to 0}{\sim} -g^3 \sum_{n=0}^{\infty} b_n g^{2n} \tag{3.22}$$

with [16–19]

$$b_0 = 11\,(4\pi)^{-2}, \qquad b_1 = 102\,(4\pi)^{-4}. \tag{3.23}$$

The 3-loop coefficient $b_2$ depends on our choice of running coupling and is presently not known.

In the following a key rôle is played by the step scaling function $\sigma(s,u)$. For any given scale factor $s$ and initial value $u = \bar{g}^2(L)$, the coupling $u' = \bar{g}^2(sL)$ may be computed by integrating the renormalization group equation (3.21) (assuming the $\beta$–function is known). $u'$ is a well-determined function of $s$ and $u$, and so we may define

$$\sigma(s,u) = u'. \tag{3.24}$$

In other words, the step scaling function is an integrated form of the $\beta$–function, which tells us what happens to the coupling if the box size is changed by a factor $s$.

Table 1. 1-loop coefficients $\delta_1(a/L)$ and $\epsilon_1(a/L)$ [eqs.(3.26),(3.28)]

| $L/a$ | $\delta_1$ | $\epsilon_1$ |
|---|---|---|
| 6 | $-0.00394$ | 0.0677 |
| 8 | $-0.00194$ | 0.0336 |
| 10 | $-0.00117$ | 0.0204 |
| 12 | $-0.00079$ | 0.0138 |
| 14 | $-0.00057$ | 0.0100 |
| 16 | $-0.00043$ | 0.0076 |

*3.6 Cutoff effects*

We finally address the question of how strongly the evolution of the coupling is affected by the underlying space-time lattice. Let us consider two lattices with size $L$ and $sL$ at the same bare coupling $g_0$ (we assume $L$ and $sL$ are integer multiples of $a$). Suppose $u$ and $u'$ are the values of $\bar{g}^2$ on these lattices and define the lattice step scaling function through

$$\Sigma(s, u, a/L) = u'. \tag{3.25}$$

Close to the continuum limit one expects that the relative deviation

$$\frac{\Sigma(2, u, a/L) - \sigma(2, u)}{\sigma(2, u)} = \delta_1(a/L)\, u + \delta_2(a/L)\, u^2 + \ldots \tag{3.26}$$

converges to zero with a rate roughly proportional to $a/L$. Since we are using an improved action, the 1-loop coefficient $\delta_1(a/L)$ is actually decreasing more rapidly (see table 1). The next term in the expansion (3.26) is not improved, however, and so, barring accidental cancellations, will be of order $a/L$.

The largest coupling at which the step scaling function will be computed is $u = 2.77$. Table 1 thus suggests that it is only weakly affected by the lattice cutoff. Whether this is really true, must of course be verified by performing numerical simulations at different values of the lattice spacing (cf. subsect. 5.1). One should also not conclude that the lattice effects are generally small as a result of our use of an improved action. A sobering example here is the quantity $\bar{v}$ defined in subsect. 3.3. On the lattice we have

$$\bar{v}(L) = \Omega(\bar{g}^2(L), a/L), \tag{3.27}$$

where $\Omega(u, a/L)$ converges to a universal function $\omega(u)$ in the continuum limit. As may be seen from table 1, the relative deviation

$$\frac{\Omega(u, a/L) - \omega(u)}{\omega(u)} = \epsilon_1(a/L) + \epsilon_2(a/L)\,u + \ldots \tag{3.28}$$

is larger in this case.

## 4. Numerical simulation

The running coupling $\bar{g}^2$ is inversely proportional to the expectation value of

$$\Delta S = \left.\frac{\partial S}{\partial \eta}\right|_{\eta=\nu=0}. \tag{4.1}$$

This observable may be computed with little effort, for any given field configuration, and so it is possible to evaluate its expectation value by numerical simulation of the Schrödinger functional. We here describe some of the technical details of this calculation and list the data obtained in this way.

*4.1 Monte Carlo algorithm*

To generate a representative ensemble of gauge field configurations, simulating the partition function (2.2) with the required boundary conditions, we use a "hybrid over-relaxation" (HOR) algorithm (for a recent review and references see ref.[20]). The basic cycle consists of 1 heatbath update sweep through the lattice followed by $N$ over-relaxation sweeps. All links are visited and updated once in every sweep. The program processes one time-slice after the other, and each time-slice is further divided into sublattices which can be updated independently of each other.

Both the heatbath and the microcanonical algorithm proceed through embedded SU(2) subgroups [21]. We use the three obvious subgroups which leave one of the basis vectors in the fundamental representation of SU(3) invariant. The situation then is essentially the same as in the SU(2) theory discussed in ref.[3]. In particular, we again employ the heatbath algorithm of Haan and Fabricius [22,23], and the exactly energy preserving moves of the embedded SU(2) links are carried out in the usual way. This is technically easier than proposing approximately microcanonical steps in the full SU(3) group followed by a non-trivial acceptance decision [24,25].

### *4.2 Autocorrelations and timing*

The number $N$ of over-relaxation sweeps per heatbath sweep is a free parameter of the HOR algorithm. In spin models and free field theory it is known that minimal autocorrelation times result, if $N$ is taken to be an approximately constant multiple of the correlation length in lattice units. In our computations the relevant scale is $L$ and $N \approx L/2a$ proves to be a nearly optimal choice. We also found it profitable to determine $\Delta S$ after each over-relaxation sweep. Little additional CPU time is required for this, while a significant loss of information is avoided when $N$ is large.

On the larger lattices most of the time is spent doing over-relaxation sweeps. At $L/a = 20$ our program achieves an average link update time of 15 $\mu$sec. This is about a factor 8 slower than the analogous SU(2) program, but since the variance of $\Delta S$ is smaller here, we only need twice as much computer time to obtain the running coupling to the same relative accuracy.

In most cases the observable $\Delta S$ is fluctuating about its mean with a nearly Gaussian distribution. Rare excursions to very small or negative values were however observed at the largest values of the renormalized coupling considered. These correspond to some long-time autocorrelations and make a reliable error estimation difficult. Inspired by the multi-canonical technique for first order phase transitions [26] the problem could be overcome by a simple modification of the sampling procedure. Further details are reported in appendix A.

### *4.3 Simulation results*

To study the evolution of the running coupling we have simulated pairs of lattices with sizes $L$ and $sL$ at the same bare coupling. The results of the computations are listed in table 2. There are 5 blocks of data, corresponding to 5 fixed values of $\bar{g}^2(L)$. In each case the bare coupling was tuned, using a reweighting technique, to obtain the desired value of $\bar{g}^2(L)$. The scale factor $s$ was set equal to 2 except for the last evolution step where $s = 3/2$.

The statistical errors were calculated by jacknife binning and checked by summing the autocorrelation function over an appropriate time interval when no reweighting was necessary. The precision in $\bar{g}^2$ given in table 2 was achieved by accumulating around 160k sweeps on the lattices with $L/a = 16$ and 200k sweeps if $L/a = 20$. On a single CRAY YMP processor this corresponds to

Table 2. Evolution of the renormalized coupling at fixed $\beta = 6/g_0^2$

| $\beta$ | $L/a$ | $\bar{g}^2(L)$ | $\bar{g}^2(sL)$ | $\bar{v}(L)$ | $\bar{v}(sL)$ |
|---|---|---|---|---|---|
| 8.7522 | 5 | 1.2430(12) | 1.4284(53) | 0.073(2) | 0.063(6) |
| 8.8997 | 6 | 1.2430(13) | 1.4270(50) | 0.067(2) | 0.053(6) |
| 9.0350 | 7 | 1.2430(15) | 1.4230(50) | 0.064(3) | 0.063(5) |
| 9.1544 | 8 | 1.2430(14) | 1.4250(58) | 0.071(2) | 0.065(6) |
| 8.1555 | 5 | 1.4300(21) | 1.6986(71) | 0.072(3) | 0.054(5) |
| 8.3124 | 6 | 1.4300(20) | 1.6859(73) | 0.069(3) | 0.064(5) |
| 8.4442 | 7 | 1.4300(18) | 1.7047(77) | 0.065(2) | 0.064(6) |
| 8.5598 | 8 | 1.4300(21) | 1.6966(90) | 0.066(2) | 0.071(8) |
| 7.5687 | 5 | 1.6950(26) | 2.101(11) | 0.070(2) | 0.066(5) |
| 7.7170 | 6 | 1.6950(26) | 2.091(11) | 0.066(2) | 0.058(5) |
| 7.8521 | 7 | 1.6950(28) | 2.112(10) | 0.064(2) | 0.052(5) |
| 7.9741 | 8 | 1.6950(28) | 2.096(11) | 0.063(2) | 0.060(5) |
| 8.1650 | 10 | 1.6950(31) | 2.092(15) | 0.066(2) | 0.053(6) |
| 6.9671 | 5 | 2.1000(41) | 2.750(17) | 0.066(2) | 0.050(5) |
| 7.1214 | 6 | 2.1000(39) | 2.783(18) | 0.059(2) | 0.047(4) |
| 7.2549 | 7 | 2.1000(43) | 2.774(16) | 0.058(2) | 0.059(4) |
| 7.3632 | 8 | 2.1000(45) | 2.772(17) | 0.059(2) | 0.059(4) |
| 7.5525 | 10 | 2.1000(42) | 2.743(24) | 0.058(2) | 0.040(6) |
| 6.5512 | 6 | 2.7700(69) | 3.489(22) | 0.053(2) | 0.047(3) |
| 6.7860 | 8 | 2.7700(73) | 3.448(28) | 0.051(2) | 0.043(3) |
| 6.9748 | 10 | 2.7700(76) | 3.487(29) | 0.052(2) | 0.045(4) |
| 7.1190 | 12 | 2.770(11) | 3.496(42) | 0.051(3) | 0.050(5) |

$s = \frac{3}{2}$ at $\bar{g}^2(L) = 2.770$ and $s = 2$ in all other cases

180 and 600 hours of CPU time respectively.

An additional set of simulations were performed to determine the bare coupling as a function of $L/a$ at fixed $\bar{g}^2(L) = 3.48$ (table 3). These results will be used in subsect. 5.3 to make contact with the physical low-energy scales of the theory.

Table 3. Bare couplings vs. lattice size at $\bar{g}^2(L) = 3.48$

| $L/a$ | $\beta$ | $L/a$ | $\beta$ |
|---|---|---|---|
| 4 | 5.9044(38) | 8 | 6.4527(46) |
| 5 | 6.0829(35) | 9 | 6.5539(80) |
| 6 | 6.2204(25) | 12 | 6.775(13) |
| 7 | 6.3443(43) | 15 | 6.973(10) |

## 5. Discussion of results

*5.1 Extrapolation to the continuum limit*

From the numbers listed in table 2 one obtains the lattice step scaling function

$$\Sigma(s, \bar{g}^2(L), a/L) = \bar{g}^2(sL) \tag{5.1}$$

at 5 values of $\bar{g}^2(L)$ and for a range of $a/L$. We now pass to the continuum limit $a/L = 0$ by extrapolating these data, using an ansatz of the form

$$\Sigma(s, \bar{g}^2, a/L) = \sigma(s, \bar{g}^2) \left\{1 + \rho(s, \bar{g}^2)\, a/L\right\}. \tag{5.2}$$

As shown by fig. 2 the fit works perfectly. Within errors there is no observable cutoff dependence, i.e. the slope $\rho(s, \bar{g}^2)$ is compatible with zero in all cases. The results of the fit will be discussed below.

Compared to the SU(2) theory the lattice step scaling function here is much less dependent on the cutoff. Presumably this is due to our use of an improved action. It is interesting to note in this connection that the SU(2) data of ref.[3] can be fitted over the whole range of couplings with an error term of the form $\rho(2, \bar{g}^2) = \rho_2 \bar{g}^2$. Moreover, the fit parameter $\rho_2$ is found to have the same sign and order of magnitude as the coefficient of the $a/L$ lattice correction which one obtains at 1-loop order of perturbation theory (recall that in the SU(2) theory the simulations were performed with the Wilson action). So it is quite plausible that the leading cutoff effects may be cancelled by a perturbative improvement of the action. Further studies are however needed to completely clarify the issue.

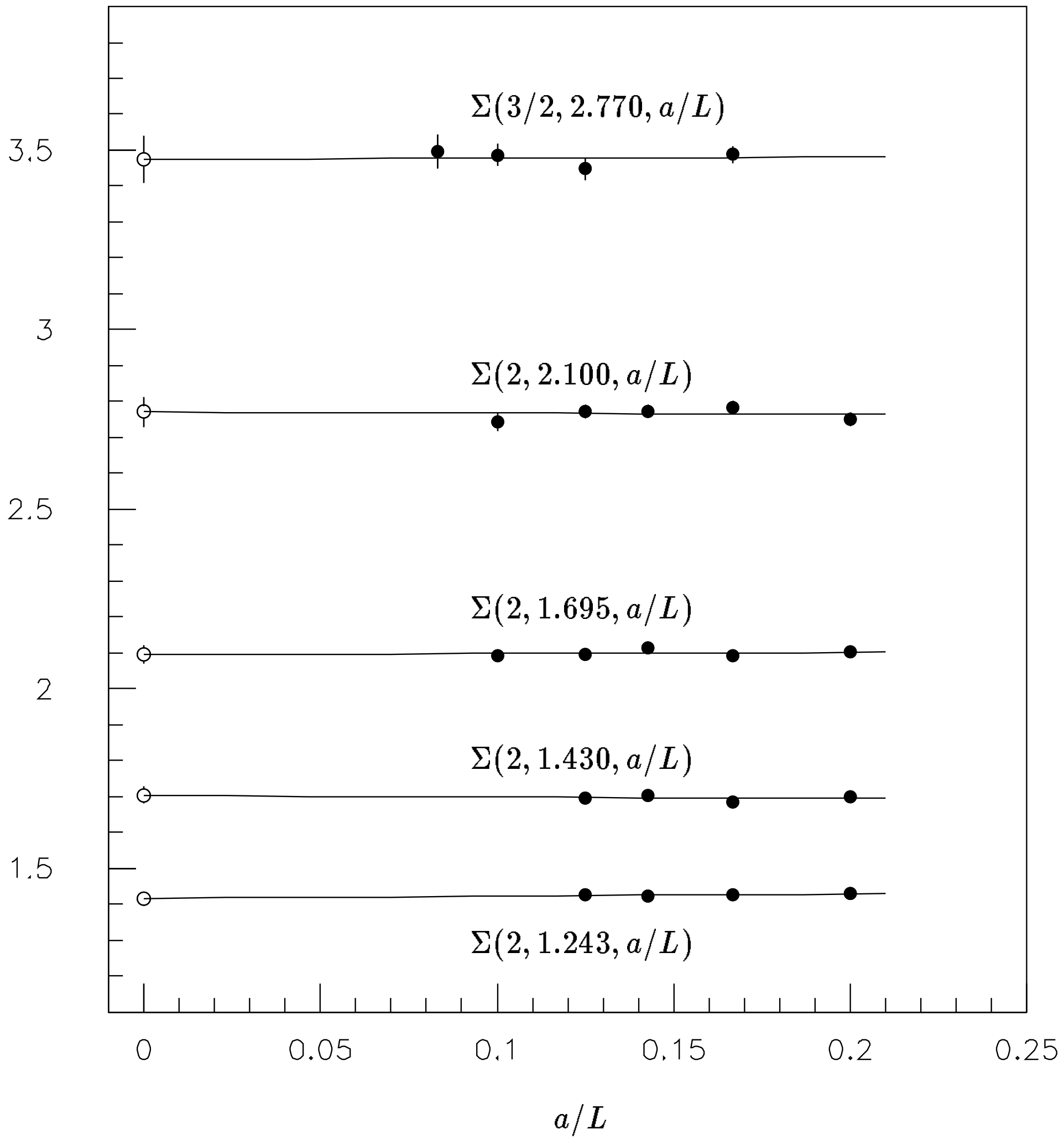

Fig. 2. Extrapolation of the lattice step scaling function to the continuum limit. The points at $a/L = 0$ represent the extrapolated values.

We now proceed to discuss the cutoff dependence of $\bar{v}$, or, more precisely, of the function $\Omega(\bar{g}^2, a/L)$ [eq.(3.27)]. The data listed in the last two columns of table 2 show that $\bar{v}$ is hardly changing as $a/L$ is made smaller at fixed $\bar{g}^2$. Note that in most cases we have results for $a/L$ from 0.2 down to 0.05 (small variations of $\bar{g}^2$ can here be ignored, because $\Omega$ is a flat function of $\bar{g}^2$). This nicely confirms the universality of the Schrödinger functional [2], although it must be said that the test is not too significant, since the statistical errors are quite large.

Table 4. Values of the step scaling function

| $\bar{g}^2$ | $s$ | $\sigma(s,\bar{g}^2)$ | $\sigma(s,\bar{g}^2)_{2-\text{loop}}$ |
|---|---|---|---|
| 1.243 | 2 | 1.416(16) | 1.428 |
| 1.430 | 2 | 1.703(24) | 1.684 |
| 1.695 | 2 | 2.095(25) | 2.071 |
| 2.100 | 2 | 2.771(41) | 2.732 |
| 2.770 | 3/2 | 3.474(65) | 3.397 |

For the extrapolation to the continuum limit, we use the formula

$$\Omega(\bar{g}^2, a/L) = \omega(\bar{g}^2)\left\{1 + \epsilon_1(a/L) + \lambda(\bar{g}^2)\, a/L\right\}, \tag{5.3}$$

where $\epsilon_1(a/L)$ is the exact 1-loop lattice correction (cf. subsect. 3.6). The remaining cutoff effects (those proportional to $\lambda$) then are compatible with zero and certainly less than 10% in the whole fit range.

*5.2 Evolution of the running coupling*

Our results for the step scaling function in the continuum limit are compiled in table 4. Up to small mismatches the couplings are arranged so that one can step down the energy scale successively, starting from the smallest coupling, $\bar{g}^2 = 1.243$, and taking a zigzag course through the table until one arrives at the largest coupling. The total scale factor then is approximately equal to 24.

Following ref.[3] the integration of the step scaling function can be performed with all errors and mismatches properly accounted for. At this point it is convenient to use $L_{\text{max}}$, the box size at which $\bar{g}^2 = 3.48$, as a reference scale. The results are listed in table 5.

It is now interesting to compare the evolution of the coupling with the predictions of perturbation theory. In the last column of table 4 we have included the values of the step scaling function which one obtains by integrating the renormalization group equation (3.21) using the 2-loop approximation to the $\beta$–function. There is hardly any difference between these numbers and the simulation results. Only at the larger couplings does one see a small sys-tematic deviation which is statistically significant, because the errors on the step scaling function are uncorrelated (they are extracted from disjoint sets of simulations).

Table 5. Running coupling at scales $L$ given in units of $L_{\max}$

| $L/L_{\max}$ | $\bar{g}^2(L)$ | $L/L_{\max}$ | $\bar{g}^2(L)$ |
|---|---|---|---|
| 1.000 | 3.480 | 0.165(9) | 1.695 |
| 0.664(19) | 2.770 | 0.084(6) | 1.430 |
| 0.332(14) | 2.100 | 0.040(4) | 1.243 |

In view of these remarks it is not surprising that a perfect representation of the data listed in table 5 can be obtained by integrating the renormalization group equation (3.21), starting from $\bar{g}^2 = 3.48$ and assuming

$$\beta(\bar{g}) = -b_0\bar{g}^3 - b_1\bar{g}^5 - b_2^{\rm eff}\bar{g}^7. \tag{5.4}$$

A correlated least squares fit yields

$$b_2^{\rm eff} = 1.5(8) \times (4\pi)^{-3} \tag{5.5}$$

with a $\chi^2$ per degree of freedom equal to 1.4/4. There is no reason to expect that $b_2^{\rm eff}$ coincides with $b_2$. But it is interesting to note that $b_2^{\rm eff}$ is about twice as large as the exact 3-loop coefficient in the minimal subtraction scheme of dimensional regularization [27]. In particular, the correction appears to be reasonably small, when the $\beta$–function is written as power series in $\alpha$.

In the range of couplings considered, there is, therefore, no discernible discrepancy between perturbation theory and the simulation results. We mention in this connection that the function $\omega(\bar{g}^2)$ has a perturbation expansion of the form

$$\omega(\bar{g}^2) = \omega_0 + \omega_1\bar{g}^2 + \ldots, \tag{5.6}$$

where the 1-loop coefficient $\omega_0$ is given by eq.(3.20). $\omega_1$ is presently not known, but the simulation data can be fitted rather well with eq.(5.6) and a reasonable effective 2-loop coefficient (see fig. 3).

We would like to emphasize, however, that our results do not prove that perturbation theory provides a good approximation to all quantities of interest up to couplings as large as 3.48. Such a general statement is bound to be false

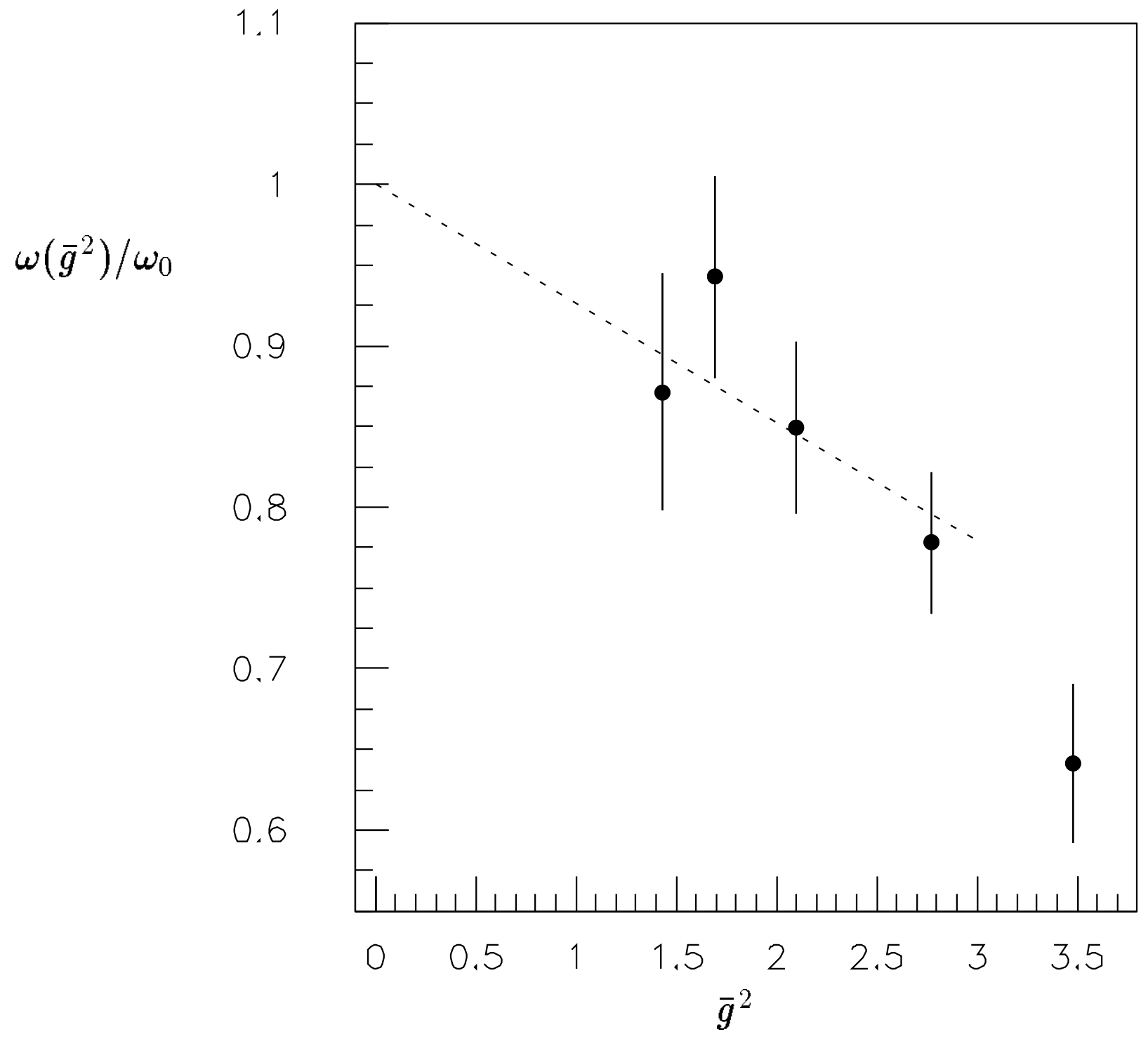

Fig. 3. Plot of $\bar{v} = \omega(\bar{g}^2)$. The dashed line is a fit, using eq.(5.6) with an adjustable 2-loop coefficient, to all data points except the one at the largest coupling.

and the running coupling in our scheme may very well turn out to be an exceptional case.

*5.3 Physical units*

We now need to relate $L_{\mathrm{max}}$ to the dynamical low-energy scales of the theory in infinite volume. As discussed in sect. 1, we decided to take the distance $r_0$ as the basic reference scale. Our first goal is thus to compute the conversion factor $L_{\mathrm{max}}/r_0$.

The heavy quark force $F(r)$ has previously been calculated through numerical simulation of the Wilson theory on large lattices. Here we use the results of refs.[8,28] to evaluate $r_0/a$ for various values of the bare coupling (see table 6). The details of the computation are as in the SU(2) theory [5]. In particular, the errors quoted include an estimate of the uncertainty which arises from a possible admixture of excited energy levels to the heavy quark

Table 6. Low-energy scales at various values of $\beta$

| $\beta$ | $r_0/a$ | $L_{\rm max}/a$ | $L_{\rm max}/r_0$ | $a^2K$ | $r_0\sqrt{K}$ | reference |
|---|---|---|---|---|---|---|
| 6.0 | 5.44(26) | 4.516(13) | 0.830(40) | 0.0513(25) | 1.23(7) | [10,28] |
| 6.2 | 7.38(25) | 5.815(11) | 0.788(27) | 0.0250(4) | 1.16(3) | [10,28] |
| 6.4 | 9.90(54) | 7.488(21) | 0.756(42) | 0.0139(4) | 1.16(6) | [10,28] |
| 6.5 | 11.23(21) | 8.498(31) | 0.756(14) | 0.0114(4) | 1.20(3) | [8] |

$F(r)$ and the value of $a^2K$ are taken from the references quoted in the last column

potential.

The values of $L_{\rm max}/a$ listed in the third column of table 6 have been obtained from table 3 using the interpolation formula $\beta = a_1 + a_2 \ln(L/a)$. It is now trivial to form the ratio $L_{\rm max}/r_0$ (4th column of table 6). $L_{\rm max}/r_0$ depends on the cutoff and must be extrapolated to the continuum limit. Assuming a linear dependence on the lattice spacing, the extrapolation yields

$$L_{\rm max}/r_0 = 0.674(50) \tag{5.7}$$

at $a/L_{\rm max} = 0$ (see fig. 5). The data are certainly consistent with this procedure, but the errors are quite large and there are only 4 data points. To be able to do better we would need a more precise determination of the heavy quark force in the range of $\beta$'s considered. One more value of $r_0/a$ at say $\beta = 6.8$ would further stabilize the extrapolation (the data of ref.[10,28] are from a physically small lattice and do not reach sufficiently large distances $r$). Such calculations are possible with the available hardware and will hopefully be done at some point.

In table 6 we have included the values of the string tension quoted in refs.[8,10]. As an additional check on the systematic uncertainty arising from the extrapolation of $F(r)$ to large distances, we have also made our own fits. In particular, the fit range was varied and an $r^{-3}$ correction besides the usual $r^{-2}$ term was included in the fit formula. Taking into account the spread of values obtained in this way, we get $r_0\sqrt{K} = 1.22(8)$, the result quoted in sect. 1.

For the purpose of illustration and to further the intuitive understanding of our results, we now convert to more physical units by setting $r_0 = 0.5\,\text{fm}$.

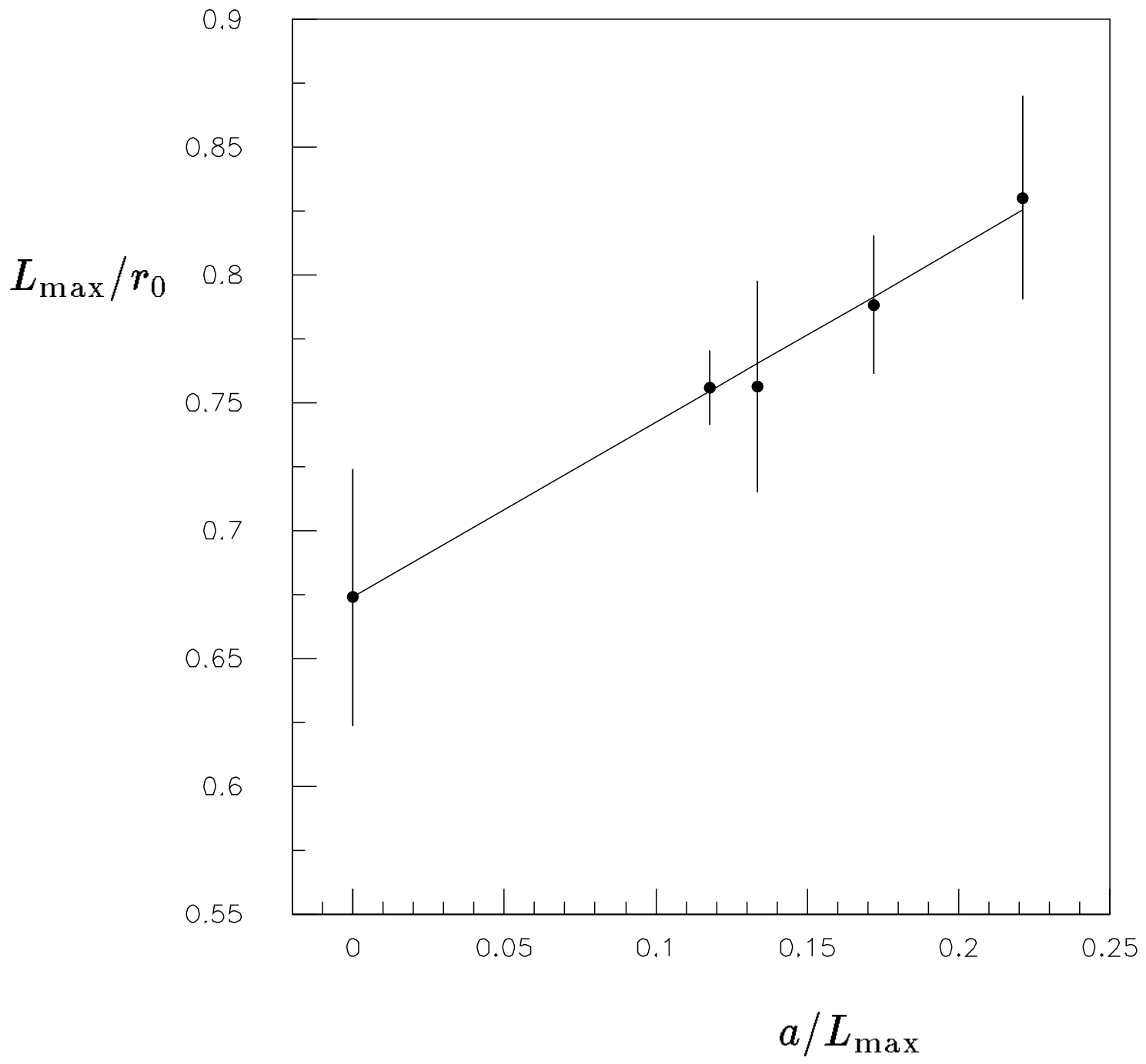

Fig. 4. Extrapolation of $L_{\rm max}/r_0$ to the continuum limit. The leftmost point represents the extrapolated value (5.7).

We then deduce that $L_{\rm max} = 0.337(25)\,{\rm fm}$, and the smallest box size covered by table 5 is hence roughly equal to 0.014 fm. For the string tension we get $\sqrt{K} = 481(32)\,{\rm MeV}$, which is somewhat larger than the values usually quoted. These are obtained from the phenomenological heavy quark potential at distances around 1 fm and also from other sources, such as the relativistic string model and the observed slope of the $\rho$ meson Regge trajectory. In the case of the heavy quark potential, the difference arises from the fact that the theoretical potential (as computed in lattice gauge theory) is steeper than the phenomenological potentials commonly used to fit the charmonium spectrum [5].

In fig. 5 the running coupling $\alpha(q)$ [eq.(3.16)] is plotted as a function of the momentum $q$ given in physical units. The error bars on the data points in this figure are barely visible. They represent the statistical errors as inferred from table 5, but not the overall scale uncertainty which originates from the conversion factor (5.7). The latter amounts to a multiplication of the energy scale by a *constant* factor and so has no bearing on the scaling properties of the coupling.

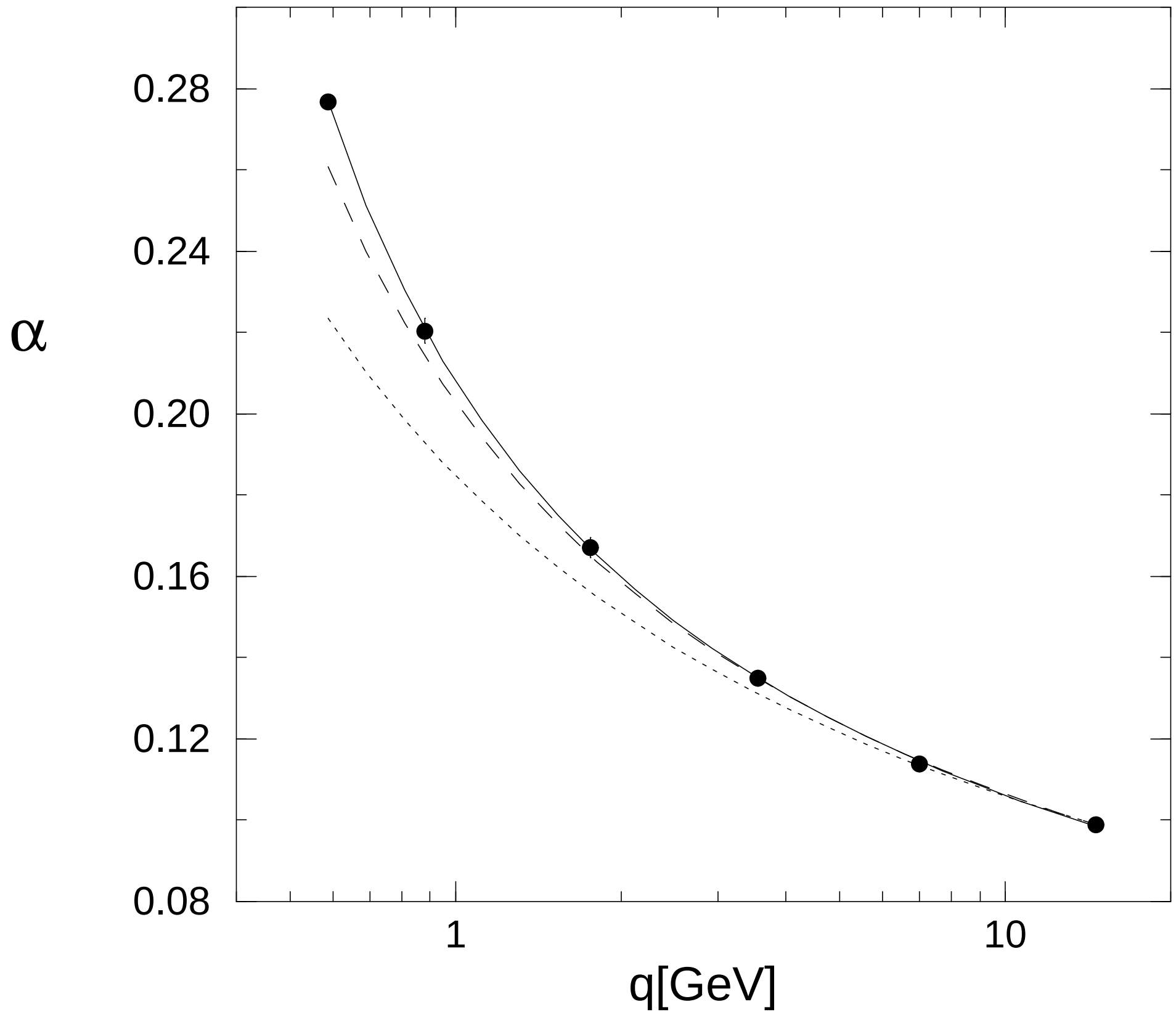

Fig. 5. Comparison of numerically computed values of the running coupling (data points) with perturbation theory. The full line represents the fit described in subsect. 5.2, while the dashed (dotted) curve is obtained by integrating eq.(3.21), starting at the right-most point and using the 2-loop (1-loop) $\beta$–function.

### *5.4 Computation of* $\alpha_{\overline{\mathrm{MS}}}$

Using the fit of our data discussed in subsect. 5.2, and the conversion factor (5.7), the running coupling $\alpha(q)$ is obtained at all momenta $q$ between $1.5 \times r_0^{-1}$ and $37 \times r_0^{-1}$. In view of fig. 5 we might actually go to even larger momenta with little risk of running into extrapolation errors.

We may now convert to the $\overline{\mathrm{MS}}$ scheme of dimensional regularization by applying the 1-loop formula (3.17). Of course this is only possible at sufficiently high energies where the coupling is small. A typical result is

$$\alpha_{\overline{\mathrm{MS}}}(q) = 0.1108(23)(10) \quad \text{at} \quad q = 37 \times r_0^{-1}, \tag{5.8}$$

where the first error is the combined statistical error from the fit and the conversion factor (5.7), while the second is equal to $\alpha(q)^3$ and thus indicates the expected order of magnitude of the 2-loop correction in the matching formula (3.17). This latter error cannot be reliably assessed until the necessary 2-loop computations have been performed. But it is reassuring to note that a correction of the quoted order of magnitude will result if $b_2$ is in the range of the effective 3-loop coefficient $b_2^{\mathrm{eff}}$ [eq.(5.5)].

### *5.5 Comparison with other computations of the running coupling*

In the work of El-Khadra et al. [6] the running coupling is obtained in two steps. One first determines the mass splitting $\Delta m$ between the 1P and 1S (quenched) charmonium levels on a large lattice, where finite volume effects can be neglected. If one takes $\Delta m$ as the low-energy reference scale, this computation may be regarded as giving the lattice spacing $a$ in physical units. The running coupling at the momentum $q = \pi/a$, which is now known in physical units, is then calculated using the "improved" 1-loop formula [29–32]

$$\alpha_{\overline{\mathrm{MS}}}(q) = \tilde{\alpha}_0/(1 + 0.309 \times \tilde{\alpha}_0). \tag{5.9}$$

The modified bare coupling $\tilde{\alpha}_0$ occurring here is defined through [29]

$$\tilde{\alpha}_0 = g_0^2/(4\pi P), \qquad P = \tfrac{1}{3}\langle \mathrm{tr}\, U(p)\rangle, \tag{5.10}$$

and the plaquette expectation value $P$ is to be computed at the given value of $g_0$.

To be able to compare with our results in a more direct way, we use $r_0$ instead of $\Delta m$ as the reference scale. From the second column of table 6 we can then read off the lattice spacing in physical units for a range of bare couplings. So if we choose $\beta = 6.5$, for example, we have $q = 35.3(7) \times r_0^{-1}$ and, noting $P = 0.6384$ [8], one obtains $\alpha_{\overline{\rm MS}}(q) = 0.1111$. Our value for the $\overline{\rm MS}$ coupling at this momentum is 0.1119(23)(10) and a similar agreement is found at the lower values of $\beta$ in table 6.

The corrections to eq.(5.9) are thus at most a few percent in this range of bare couplings. Note that if one employs another reference scale, such as $\Delta m$, the statement remains true provided the corresponding conversion factor can be shown to be independent of the lattice spacing, to a sufficient degree of precison, at the values of $\beta$ considered.

In refs.[7–11] the coupling $\alpha_{\rm q\bar{q}}(q)$ [eq.(3.18)] is computed from lattice data on the heavy quark force. A subtle fit and subtraction of cutoff effects is needed at short distances, where the raw data are strongly dependent on the lattice action employed. The values of $\alpha_{\rm q\bar{q}}$ obtained in this way are significantly larger than those one gets by combining the 1-loop formula (3.19) with our results for $\alpha$. At $\beta = 6.5$ and $r/a = 2.5322$, for example, the value quoted in ref.[8] is 0.248(2)(1), while our result is 0.205(7)(5). There is no reason to be worried about this discrepancy, because the distances at which the comparison can be made are rather large in physical units. So it could well be that the non-perturbative corrections to the matching formula (3.19) are not negligible. It is also not evident to us that the cutoff effects are indeed under control, when $\alpha_{\rm q\bar{q}}$ is determined from the heavy quark potential at distances as low as 2 or 3 lattice spacings.

## 6. Concluding remarks

Compared to our earlier study of the SU(2) Yang-Mills theory [3], the results obtained here are very similar. In particular, we again find that the perturbative scaling regime and the low-energy domain of the theory are smoothly connected, with no complicated transition region. We have also been able to determine $\alpha_{\overline{\rm MS}}(q)$ at momenta $q$ up to about 14 GeV, with an estimated precison of a few percent. An interesting observation is that the relation

$$2\alpha_{\overline{\rm MS}}(q)\big|_{\rm SU(2)} = 3\alpha_{\overline{\rm MS}}(q)\big|_{\rm SU(3)} \tag{6.1}$$

holds within errors, if $q$ is given in units of $r_0$. Eq.(6.1) is consistent with the evolution of the coupling up to 3-loop order of perturbation theory and suggests that $N\alpha_{\overline{\rm MS}}$ is only weakly dependent on the number $N$ of colours.

A critical step in our computations is the evaluation of the conversion factor $L_{\rm max}/r_0$, which is needed to make contact with the low-energy physics of the theory in infinite volume. The situation is more difficult in the SU(3) theory than in the SU(2) case, because the large volume lattices which one has been able to simulate to date have significantly larger lattice spacings. Moreover, the statistical errors on the heavy quark force at intermediate distances (where $r_0$ is determined) are far from negligible. We hope that these deficits can be removed in the near future.

The finite-size scaling technique employed in this paper is expected to be useful in QCD, too. Besides the running coupling one is here also interested to determine the renormalized quark masses. The technical difficulties are considerable, however, and it may take a number of years until a plot like fig. 5 can be drawn for QCD with a non-zero number of light quarks. In particular, the renormalizability of the Schrödinger functional must be discussed and the performance of numerical simulation algorithms for dynamical fermions on physically small lattices must be evaluated.

We are indebted to A. Kronfeld for correspondence and to G. Bali for communicating simulation results on the heavy quark force prior to publication. The computations have been performed on the CRAY computers at HLRZ and CERN. We thank these institutions for their support.

## Appendix A

At the largest couplings considered ($\bar{g}^2 \simeq 3.48$), a straightforward simulation of the Schrödinger functional is unsatisfactory, because some rare fluctuations of the observable $\Delta S$ lead to uncomfortably large autocorrelation times. We here describe how one can get around this problem by using a variant of the multi-canonical technique.

The history and distribution of $\Delta S$ for a "normal" simulation run are shown in figs. 6a and 6b. The corresponding data for a run at $\bar{g}^2 = 3.45$ are displayed in figs. 6c and 6d (cf. table 2). In both cases we plotted block averages from 81 successive sweeps as individual "events" to smooth out the short time noise. To a good approximation they can be treated as statistically independent in normal runs. Spikes as the one clearly seen in fig. 6c however lead to longer autocorrelation times. They are associated with a long tail of the distribution of $\Delta S$ at low values of $\Delta S$ (cf. fig. 6d). The problem is that the spikes influence the expectation value of $\Delta S$ significantly, but are not sufficiently frequent for a reliable determination of the statistical error.

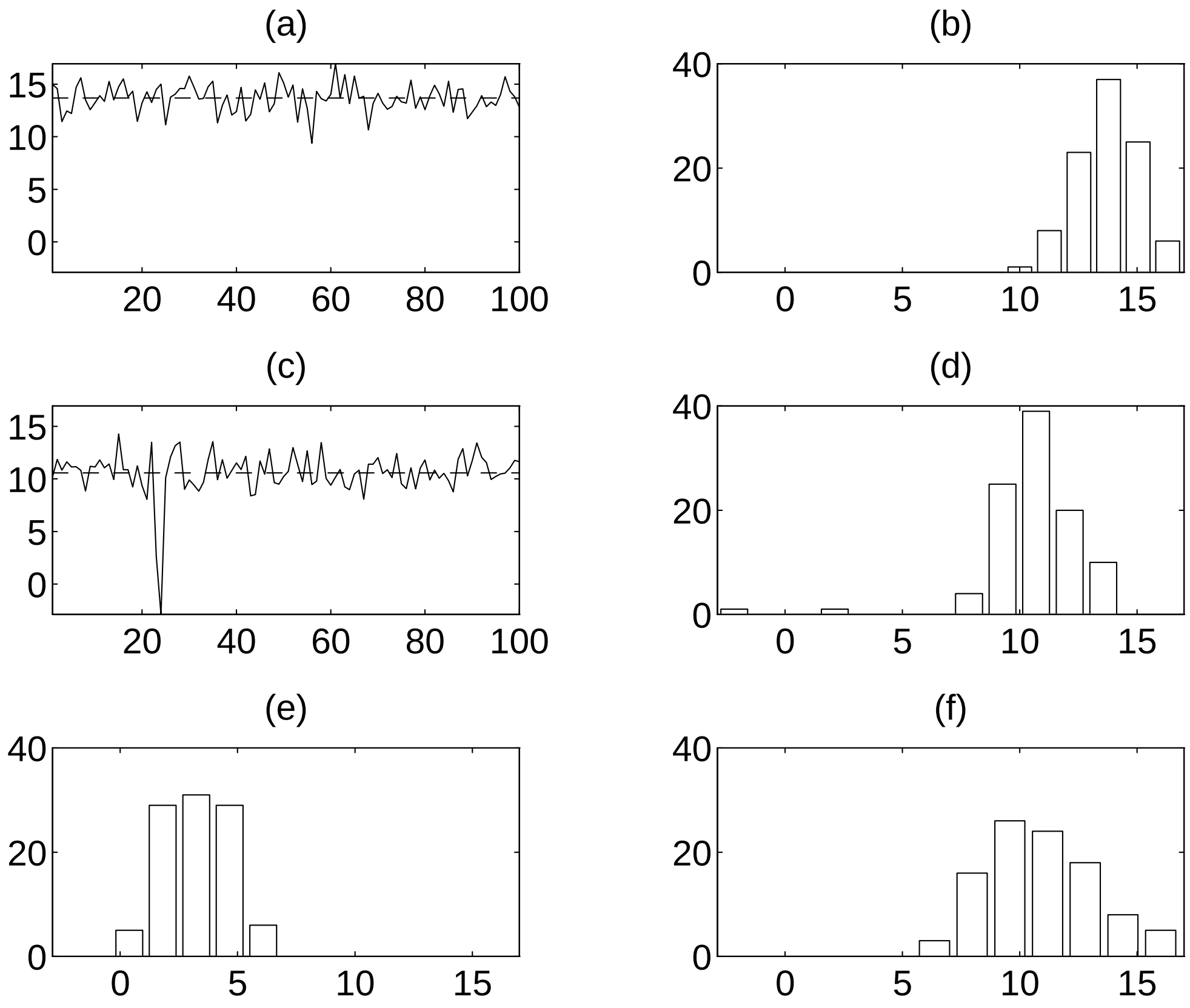

Fig. 6. Histories and distributions of $\Delta S$ at $L/a = 12$. a,b: $\beta = 7.1214$; c,d: $\beta = 6.7860$; e: the same, with modified sampling ($\gamma = 0.05$), and f: after reweighting.

The statistics in the tail of the distribution of $\Delta S$ can be enhanced by sampling the gauge fields with the modified Boltzmann weight

$$\widetilde{P}[U] \propto \exp(-S[U] - \gamma \Delta S[U]). \tag{A.1}$$

The extra term proportional to $\gamma$ is easily accounted for in the simulation algorithm. For $\gamma > 0$ the simulation is now biased towards smaller values of

$\Delta S$ and the desired expectation value is obtained through

$$\langle \Delta S \rangle = \frac{\langle \Delta S \exp(\gamma \Delta S) \rangle_{\widetilde{P}}}{\langle \exp(\gamma \Delta S) \rangle_{\widetilde{P}}}. \tag{A.2}$$

The optimal value of $\gamma$ depends on the observed width of the distribution of $\Delta S$ and requires some tuning. For $\gamma = 0.05$ the distributions of $\Delta S$ and $\Delta S \exp(\gamma \Delta S) / \langle \exp(\gamma \Delta S) \rangle_{\widetilde{P}}$ in the $\widetilde{P}$–ensemble are as shown in figs. 6e and 6f. Evidently the problem has been overcome in this way.